\UseRawInputEncoding
\documentclass[pdftex,twocolumn,epjc3]{svjour3}         
\RequirePackage[T1]{fontenc}
\smartqed  
\RequirePackage{graphicx}
\RequirePackage{placeins}
\RequirePackage{flushend}
\RequirePackage{subcaption} 
\RequirePackage[numbers, square, comma, sort&compress]{natbib} 
\RequirePackage[colorlinks,citecolor=blue,urlcolor=blue,linkcolor=blue]{hyperref}
\usepackage{amsmath}
\journalname{Eur. Phys. J. A}
\begin{document}
\title{Geometric Poisson distribution of photons produced in the ultrarelativistic hadronic collisions}
\author{Rahul R Nair\thanksref{e1,addr1}
        \and
        Grzegorz Wilk\thanksref{e2,addr1} 
        \and
        Zbigniew W\l odarczyk\thanksref{e3,addr2} 
}
\thankstext{e1}{e-mail: rrn.phy@gmail.com}
\thankstext{e2}{e-mail: grzegorz.wilk@ncbj.gov.pl}
\thankstext{e3}{e-mail: zbigniew.wlodarczyk@ujk.edu.pl}
\institute{National Centre For Nuclear Research, Pasteura 7, Warsaw 02-093, Poland.\label{addr1}\and Institute of Physics, Jan Kochanowski University, 25-406 Kielce, Poland. \label{addr2} 
}
\date{Received: date / Accepted: date}
\maketitle
\begin{abstract}
We show that the multiplicity distribution of photons produced with enhanced void probability in inelastic proton-proton collisions at $\sqrt{s} =$ 900 GeV, 2.76 TeV, and 7 TeV, measured at forward rapidities by the ALICE experiment at LHC, can be described by the geometric Poisson distribution. The traditionally used negative binomial distribution fails to reproduce the enhanced void probability and the shape of the modified combinants simultaneously. Our findings are relevant for the theoretical modeling of photon production processes in high-energy hadronic collisions.
\end{abstract}
\section{Introduction}\label{sec:intro}
Some time ago, the ALICE experiment at the LHC provided measurements of inclusive photon multiplicity distributions at various center-of-mass energies for proton-proton collisions in the forward rapidity region \cite{Abelev2015}. It turns out that the measured distributions are well described by the NBD distribution, but only if the measured void probability, $P(0)$, is not included in the fit. This means that some information hidden in the measured probability distributions, particularly the observation that $P(0) > P(1)$, is not taken into account. On the other hand, as shown in \cite{Chau:1992uq, PhysRevLett.70.3380, Huang:1996az, PhysRevD.37.2446, Huang:1997bra}, it could provide valuable insights into the physical processes leading to the kind of final state fluctuations one would measure. 

Hence, the NBD does not seem to be the right choice here, and we should consider other options. The observed particles are photons; therefore, (notwithstanding the fact that the inclusive photons produced in hadronic interactions at high energies are dominated by $\pi^0$ decays), we can treat them as a roughly degenerate gas of bosons. In such a case, we can expect to have fluctuations in the photon multiplicity as a consequence of quantum statistics \cite{Chau:1992uq}. Consider each event in which multiple photons are produced as a statistical ensemble. Assume that the identical photons produced in each event occupy a single cell in phase space. The argument that identical particles produced in a limited spatial volume and within a certain spread of momentum lie in a single elementary cell in phase space was rephrased in quantum mechanical terms in \cite{PhysRevD.10.65}.
 
For such processes, the most promising candidate for $P(n)$ is the geometric Poisson distribution (GPD), which has been previously proposed to study enhanced void probability distributions in various physical systems through various approaches to its physical origin, instead of the traditionally used negative binomial distribution (NBD) \cite{Chau:1992uq, PhysRevLett.70.3380, Huang:1996az}. It has been particularly successful in describing the probability distributions of particle production from clusters. For example, a two-stage production mechanism can be traced back to P\'olya's cluster model \cite{PhysRevD.37.2446}. If the particles obey Bose-Einstein statistics, then the probability of observing a particle in a particular cell is given by the geometric distribution. For a fixed number of cells, we have a Negative Binomial Distribution (NBD) for the number of produced particles, while for the number of cells fluctuating according to a Poisson distribution, we can expect a P\'olya-Aeppli (geometric Poisson) distribution for the number of produced particles.
 
In this paper, we analyze the photon multiplicity distributions measured by the ALICE experiment in the forward rapidity region at the LHC within the picture of Poisson-distributed emitting sources for photons. To this end, we will use the modified combinants  \cite{Kauffmann_1978, Vasudevan_1984, BALANTEKIN1991231, HEGYI1999126, Wilk_2017, Ang2020, PhysRevD.99.094045} of the multiplicity distribution as a statistical tool to determine the correct form of distribution for the ALICE photons that works well for collisions at all available center-of-mass energies. These modified combinants have been found to exhibit intriguing features that can provide valuable insights into the production process of particles in high-energy particle collisions. 

The multiplicity distribution can be characterized by the recursive formula:
\begin{equation}\label{eq2}
    (n+1)P(n+1) = g(n)P(n)
\end{equation}
where the function $g(n)$ determines the algebraic structure of $P(n)$. 
For Poisson distribution $g(n)= const$ while for NBD, $g(n)$ is increasing linear function of multiplicity $g(n)=a+bn$. By considering that $g(n)$ should reflect the interconnections between multiplicity $n$ and all lower multiplicities, Eq.~(\ref{eq2}) can be written as follows
\begin{equation}\label{eq3}
    (n+1)P(n+1) = \left<n\right>\sum_{j=0}^{n}C_jP(n-j)
\end{equation}
where the coefficients $C_j$ are the modified combinants. By reverting the recurrence relation in Eq.~(\ref{eq3}), an expression for the modified combinants can be obtained as
\begin{equation}
\label{Comb}
    \left<n\right>C_j = (j+1) \frac{P(j+1)}{P(0)}  - \left<n\right>\sum_{i=0}^{j-1}C_i \frac{P(j-i)}{P(0)} 
\end{equation}
where $\left<n\right>$ represents the mean of the multiplicity distribution. The formula in Eq.~(\ref{Comb}) allows us to obtain modified combinants $C_j$ from both the measured multiplicity distribution $P(n)$ (provided that $P(0)$ is measured or can be reliably obtained from the extrapolation to $n=0$) and from the phenomenological distribution $P(n)$ chosen to describe the experimental data \footnote{Thus the modified combinants, $C_i$, serve as a valuable additional tool to assess whether the phenomenologically selected $P(n)$ accurately describes the experimentally measured $P(n)$ or not.}. In general, they can be expressed in terms of the generating functions $G(z)$ as follows:
\begin{equation}
\label{gencj}
    \left<n\right>C_j =  \frac{1}{j!} \frac{d^{j+1}}{dz^{j+1}}\ln G(z) \big|_{z=0}
\end{equation}
where $G(z)$ is related to the multiplicity distribution $P(n)$ as follows
\begin{equation}
    G(z) = \sum^{\infty}_{n=0} P(n)z^n
\end{equation}
Our analysis, presented in this paper, reveals that the geometric Poisson distributions can accurately describe the multiplicity distributions measured by the detector for all available center-of-mass energies. Furthermore, we observe that the modified combinants extracted from the experimental photon multiplicity distribution can be well reproduced by the modified combinants obtained from the  geometric Poisson distribution  with parameter values that describe the corresponding $P(n)$. Based on these findings, we conclude that the production process of inclusive photons in proton-proton collisions results in a geometric Poisson distribution in the final state.\par
We have organized the paper as follows: in the next section, we provide a brief discussion of the experimental aspects of the data used in our analysis. In the subsequent sections, we discuss the properties of the geometric Poisson distribution and present the results of the fit with the ALICE data. Finally, we conclude the paper with a comprehensive discussion of our observations and their implications.

\section{Brief description of the experimental data}\label{sec:ALICE}
The ALICE collaboration at LHC \cite{Collaboration_2008} measured the multiplicity and pseudorapidity ($\eta$) distributions of inclusive photons in proton-proton (pp) collisions at center-of-mass energies $\sqrt{s}=$ 0.9, 2.76, and 7 TeV with a magnetic field of 0.5 T \cite{Abelev2015}. The data used in our analysis includes 2 million, 8 million, and 9 million analyzed events for the respective collision energies. The minimum bias data used for the inelastic (INEL) events were collected with the requirement of at least one hit in the ALICE Silicon Pixel Detector (SPD) or in either of the two V0 detector arrays in ALICE \cite{Aamodt_2010}. The decommissioned Photon Multiplicity Detector (PMD) of the ALICE experiment was primarily used for the event-by-event measurement of photon multiplicity distribution \cite{ALICE:PMD1,ALICE:PMD2}. The PMD, which utilized the pre-shower photon measurement technique, was located 367 cm away from the interaction point in ALICE and had an acceptance range of $2.3 <\eta < 3.9$ with full azimuthal coverage. A photon passing through the converter plate of the PMD produced an electromagnetic shower in the pre-shower plane, leading to a large signal spread over several detector cells, while the signal produced from a charged particle was mainly confined to a single cell. The differences in responses of charged particles and photons were utilized to reject charged tracks in the ALICE photon measurements. The ALICE collaboration performed corrections for detector effects based on simulated events using various tunes of PYTHIA and PHOJET event generators \cite{Abelev2015}. The corrected multiplicity distributions from these measurements are used for the calculations in this work. In the next section, we briefly revisit the Geometric Poisson (P\'olya-Aeppli) distribution, which will be used to describe the ALICE data in the subsequent sections. 
\section{Geometric Poisson (P\'olya-Aeppli) distribution}\label{sec:GP}
The Geometric Poisson or P\'olya-Aeppli distribution, introduced by George P\'olya is credited to the work of his student Alfred Aeppli in the 1920s and 1930s \cite{AIHP_1930__1_2_117_0,Aeppli}. It was named the geometric Poisson distribution by Sherbrooke in 1968 \cite{https://doi.org/10.1002/nav.3800150206}. The geometric Poisson distribution is a compound Poisson distribution that can describe objects that come in clusters, such that the number of clusters follows a Poisson distribution and the number of objects within a cluster follows a geometric distribution. If $N$ is a P\'olya-Aeppli random variable, it can be written as:
\begin{equation}
    N = \sum_{i=1}^{M} K_{i}
\end{equation}
where $M$ is a Poisson random variable with parameter $\lambda$.  The probability mass function of $M$ is given by
\begin{equation}
    P_M(m) = \frac{e^{-\lambda}\lambda^m}{m!}\;\; ; m =0,1,2...
\end{equation}
and the $K_{i}$ are identically and independently distributed shifted geometric random variables with the following common probability mass function 
\begin{equation}\label{geometric}
    P_K(k) = \theta(1-\theta)^{k-1}\;\; ; k = 1,2...
\end{equation}
where $\theta$ is a parameter. The probability mass function of a random variable $N$ distributed according to the geometric Poisson distribution is then given by \cite{doi:10.1080/00949650802711925} 
\begin{equation}\label{pmf}
    P(n,\lambda,\theta) = 
             \begin{cases} 
		e^{-\lambda} \;\;;\;n = 0 \\
		 \sum_{k = 1}^{n} \frac{e^{-\lambda}\lambda^k}{k!} \binom{n - 1}{k - 1} \theta^{k} (1 - \theta)^{n - k} \;\;;\;n\geq 1.
		\end{cases}			
\end{equation}
The mean ($\mu$) and variance($\sigma^2$) of the above distribution are
\begin{equation}
\mu =  \frac{\lambda}{\theta}, \;\; \sigma^2 = \lambda \frac{(2-\theta)}{\theta^2}.  
\end{equation}
When $\theta = 1$, the distribution reduces to the Poisson distribution with mean given by $\lambda$.  For geometric Poisson compound distribution, generating function $G(z)_{GP}$ can be writted as
\begin{equation}\label{gpgen}
    G(z)_{GP} = \exp\left( \frac{\lambda(z-1)}{1-z(1-\theta)} \right)
\end{equation}
Using Eq.~(\ref{gpgen}) in Eq.~(\ref{gencj}), we can obtain the analytic expression for $C_j$ of geometric Poisson distribution as
\begin{equation}
\label{gpcj}
    \left<n\right>C_j =  (j+1) \lambda\theta(1-\theta)^{j}
\end{equation}
For Poisson distribution $C_j = \delta_0j$ and for Negative binomial distribution $C_j=(1-p)p^j$ (where $p<1$ is the parameter). The latter decreases monotonically with increasing the rank j. There are alternative expressions for the geometric Poisson distribution in the literature, such as those involving confluent hypergeometric functions \cite{EVANS1953,doi:10.1080/03461238.1960.10410586} and Laguerre polynomials \cite{doi:10.1287/opre.7.3.362}. In the next section, we present our attempt to describe the ALICE photon multiplicity data mentioned in the previous section using the geometric Poisson distribution.

\section{Multiplicity distribution and modified combinants of ALICE inclusive photons}\label{sec:MulComb} 
The multiplicity distribution $P(n)$ of inclusively produced photons in inelastic proton-proton collisions at ALICE has been published in \cite{Abelev2015}. We fit the $P(n)$ distributions for the three center-of-mass energies using Eq.~(\ref{pmf}). 
\begin{figure}
\caption{Multiplicity distributions of inclusive photons measured in inelastic pp collisions at ALICE \cite{Abelev2015} in the $2.3 <\eta < 3.9$ interval (Filled squares) fitted with the Geometric Poisson distribution (Open squares) with parameter values given in Table~\ref{tabpara}.}\label{Multiplicities}
\includegraphics[width=0.5\textwidth]{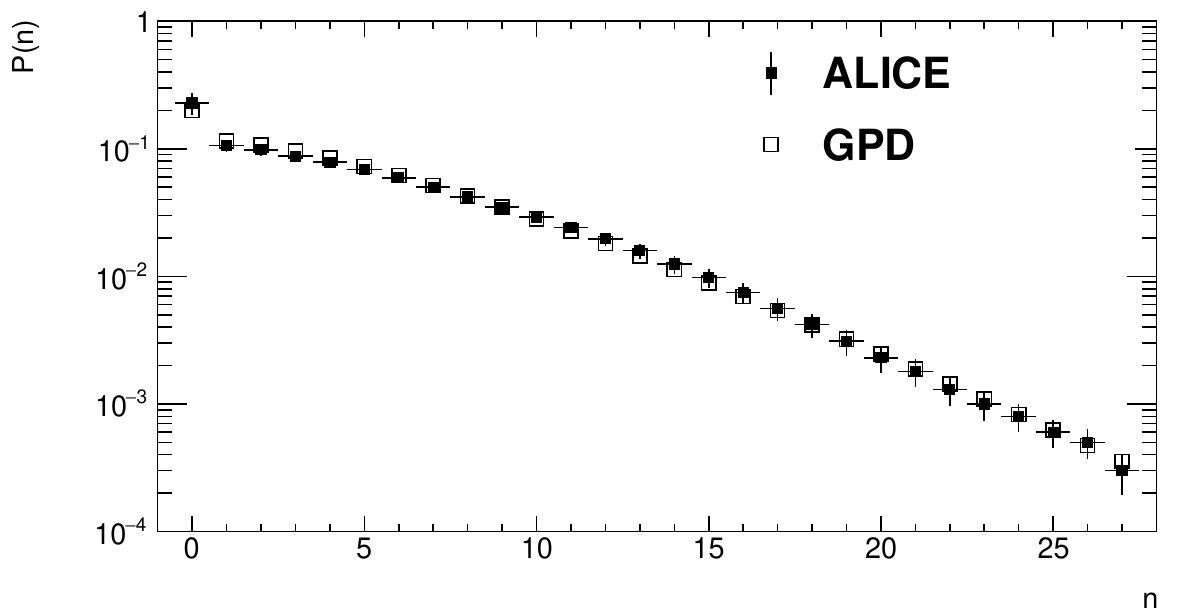}\subcaption{$\sqrt{s} = 900$ GeV}
\includegraphics[width=0.5\textwidth]{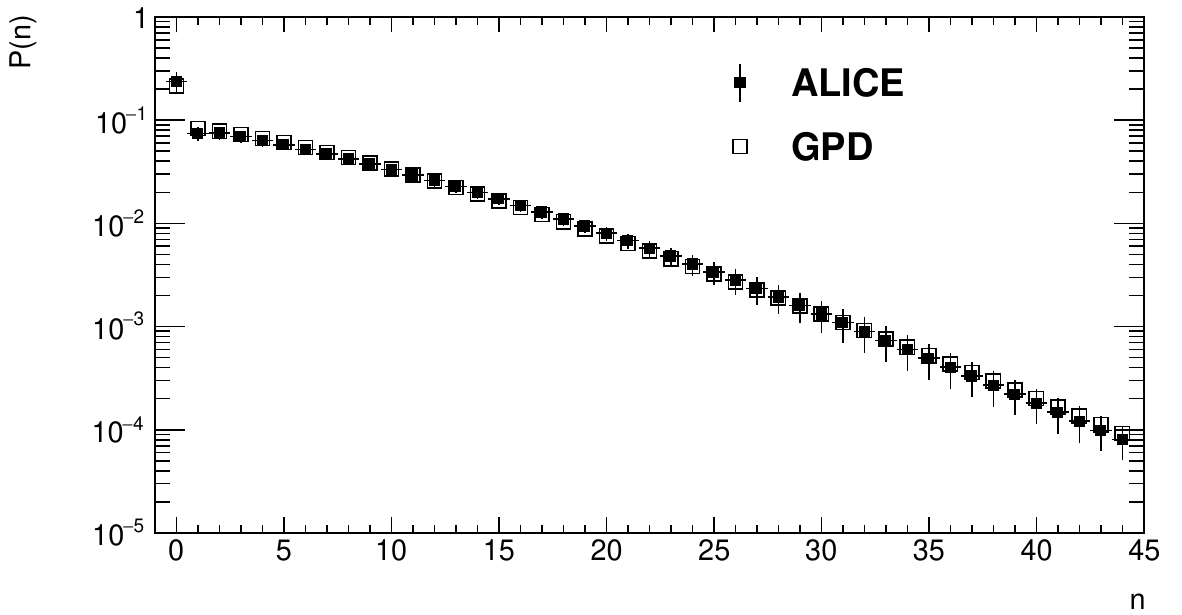}\subcaption{$\sqrt{s} = 2.76$ TeV}
\includegraphics[width=0.5\textwidth]{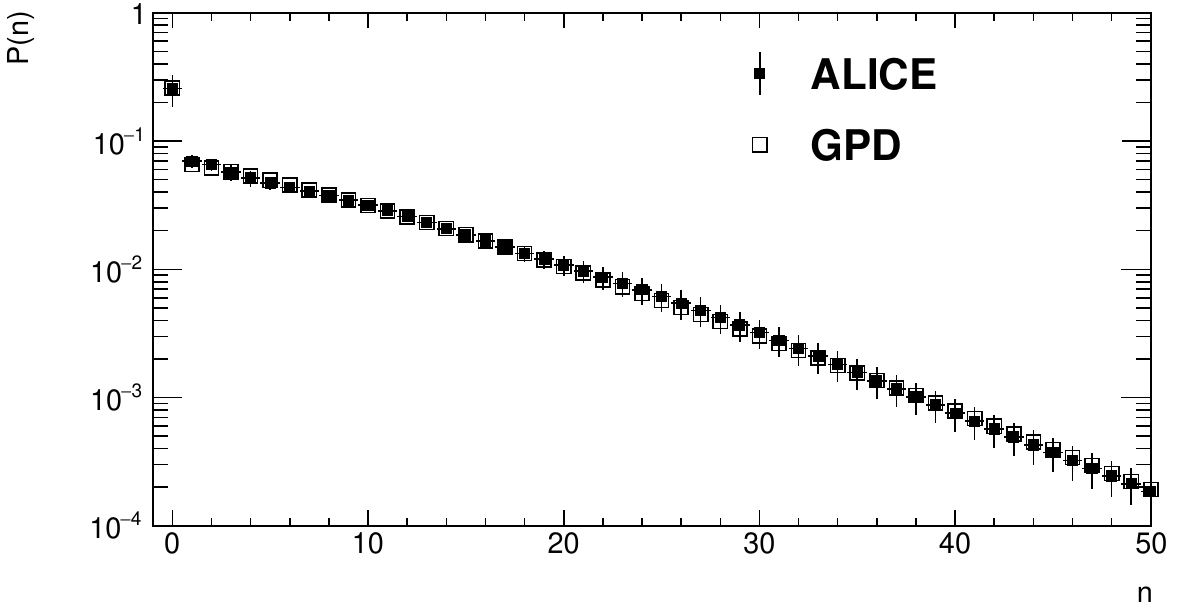}\subcaption{$\sqrt{s} = 7$ TeV}
\end{figure}
We find that the geometric Poisson distribution in the form of Eq.~(\ref{pmf}) can successfully describe the experimental $P(n)$ distribution. The results are shown in Fig. \ref{Multiplicities}, and the parameters obtained from the fit are presented in Table~\ref{tabpara}.
\begin{figure}
\caption{Modified combinants of ALICE inclusive photons (filled circles) estimated from the $P(n)$ distributions  in Fig.~\ref{Multiplicities} compared with those obtained from the Geometric Poisson distribution(open squares) Eq.~(\ref{gpcj}) with parameter values given in Table~\ref{tabpara} and with the NBD (open circles) using parameters from \cite{Abelev2015}. The solid lines are to guide the eye.}\label{Cjs}
         \includegraphics[width=0.5\textwidth]{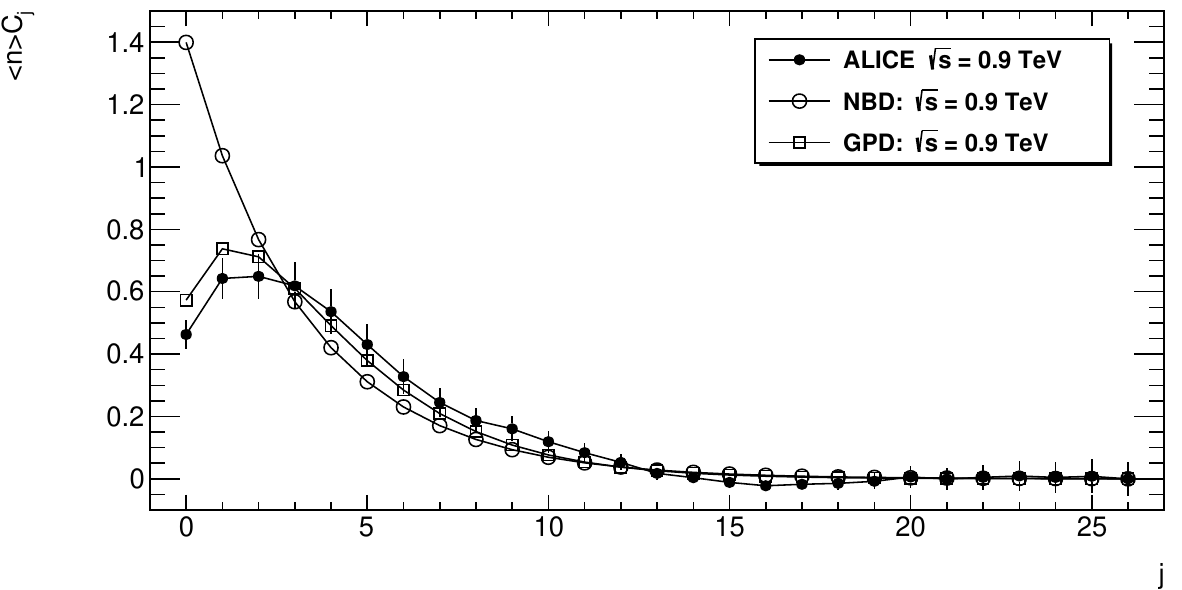}\subcaption{$\sqrt{s} = 900$ GeV}
         \includegraphics[width=0.5\textwidth]{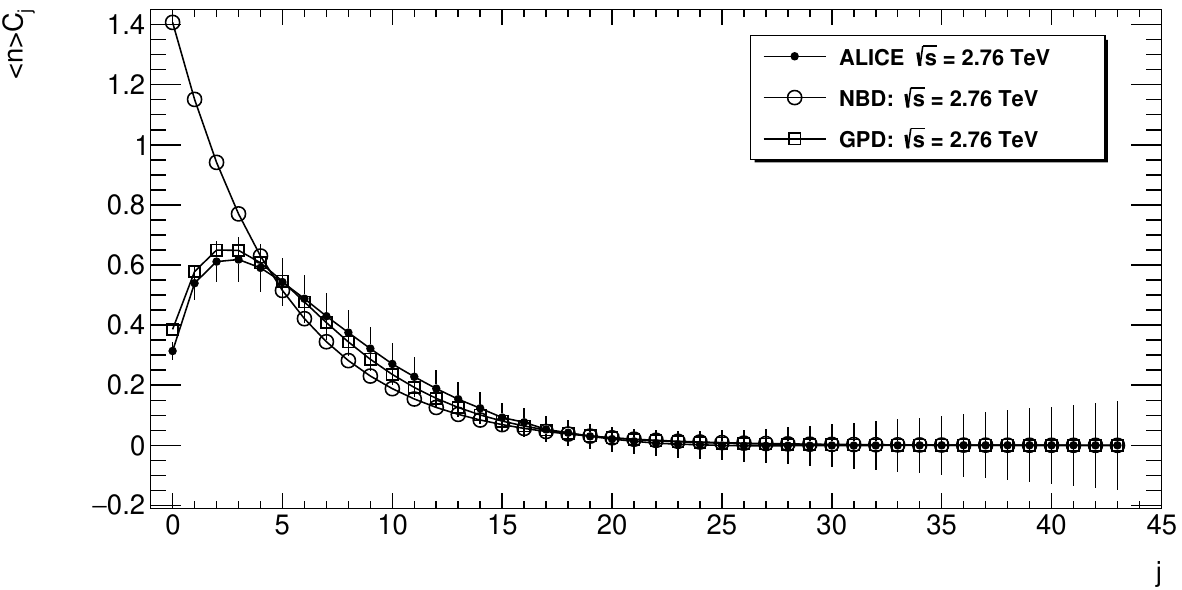}\subcaption{$\sqrt{s} = 2.76$ TeV}
         \includegraphics[width=0.5\textwidth]{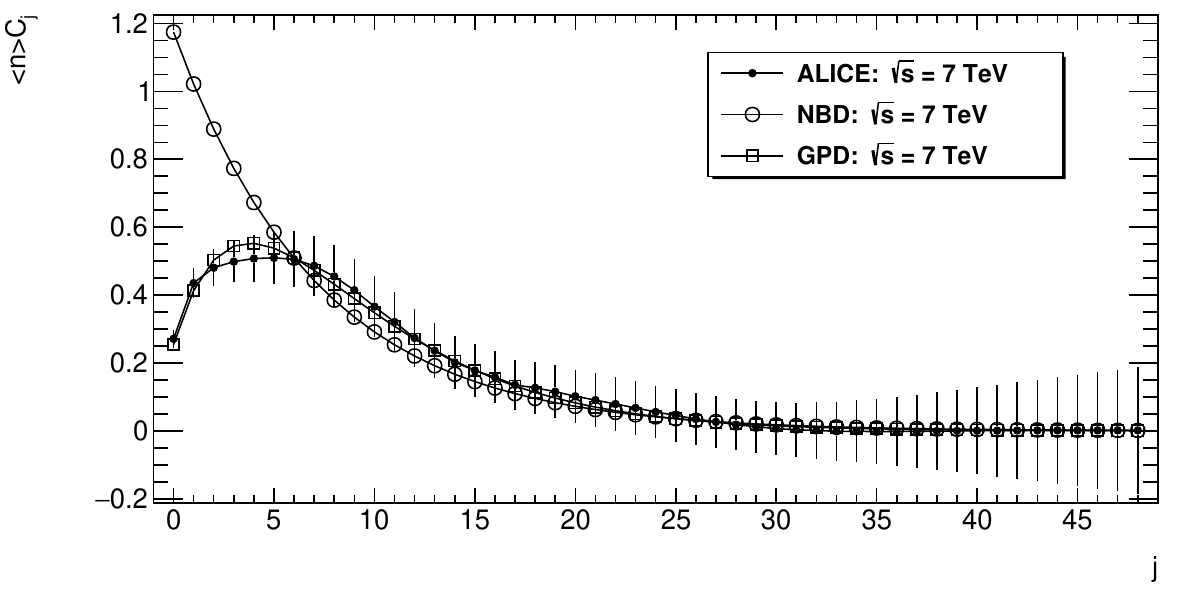}\subcaption{$\sqrt{s} = 7$ TeV}
\end{figure}
The ALICE collaboration also performed a negative binomial fit of the $P(n)$ distributions for the three collision energies in \cite{Abelev2015}, but for $n>0$. The modified combinants $C_j$ of photon multiplicity distributions are calculated using Eq.~(\ref{gpcj}) with the corresponding parameters from Table~\ref{tabpara}. We observe that the $C_j$ thus estimated from the geometric Poisson distribution can approximately reproduce the shape of the $C_j$ obtained from the experimental data. The modified combinants from NBD show differences from the experimental data for small values of $j$. This observation mostly corresponds to the differences in the void probability $P(0)$ given by 
\begin{equation}
   P(0) = \exp\left(-\sum_{j=0}^{\infty} \frac{\left<n\right>}{(j+1)}C_j\right)
\end{equation}
from the two distributions. As in the comparison demonstrated in Fig.~\ref{Cjs}, the between $C_j$s estimated from the NBD fit in \cite{Abelev2015} and our GPD description show that the latter better describes the experimental data \footnote{The recurrence formula in Eq.~(\ref{Comb}) derives $C_j$ as a recursive sum of $C_iP(j-i)$ over the index $i<j$, which leads to an increase in the uncertainties of $C_j$ of higher rank. A detailed discussion on the estimation of uncertainties of modified combinants is given in \cite{PhysRevD.99.094045, EPJC(2018)78:816}. According to the results presented therein, the errors in estimating $C_j$ in our case increase with rank $j$. However, they remain within limits that do not significantly interfere with the conclusions drawn here.}. Therefore, we argue that the geometric Poisson distribution provides a more complete description of the inclusive photon production in inelastic proton-proton collisions in the forward rapidity regions at ALICE. This observation allows us to revisit the theoretical models of photon production in high-energy hadronic collisions in a more detailed manner.
\begin{table}
\caption{Parameters of the Geometric Poisson fit to the inclusively produced photon multiplicity distributions in the pseudo rapidity interval of $2.3 <\eta < 3.9$ at ALICE.}
\begin{tabular}{llll}
\hline\noalign{\smallskip}
$\sqrt{s}$ (TeV) & $\lambda$ & $\theta$& $\chi^2/n.d.f$\\
\noalign{\smallskip}\hline\noalign{\smallskip}
0.90& $1.607\pm0.095$ &$0.357\pm0.012$&2.875/26\\
2.76& $1.538\pm0.066$ &$0.251\pm0.006$&3.329/43\\
7.00& $1.352\pm0.052$ &$0.188\pm0.004$&1.636/49\\ 
\noalign{\smallskip}\hline
\end{tabular}
\label{tabpara}
\end{table}
\section{Summary and Outlook}\label{sec:Sum} 
In the previous sections, we revisited the method of modified combinants-based analysis, the ALICE experimental photon multiplicity data in the forward rapidity regions, and some basic aspects of the geometric Poisson distribution. We observed that the measured multiplicity distributions can be well described by the geometric Poisson distribution (GPD). We also showed that the GPD can approximately reproduce the modified combinants of multiplicity estimated from the data. Based on these observations, we argue that the underlying distribution of inclusive photon production must result in a GPD in the final state.\par As mentioned earlier, various kinds of mechanisms can infact lead to a GPD in the final state. Similarly, the enhanced void probability can also be understood from different approaches such as final-state interactions, branching process approach, and two-stage production process \cite{Huang:1997bra}. In the branching-process approach, the $P(n)$ is determined by the time-evolved branching equation, where the process of a parent cluster branching into two daughter particles in the next time step has to be significant to attain the enhanced void probability.\par
However, in the two-stage mechanism with production and decay, the GPD is taken as the composition of two distributions. The clusters produced must decay, which means that the $P(0)$ for decay must be zero. Hence, the observed $P(0)$ in the multiplicity distribution must come from the production process, which is Poissonian. To achieve $P(0) > P(1)$, the processes of producing one cluster and then decaying into two particles must be significant to have the void probability enhancement in this approach. In the scenario of compound Poisson distribution, replacing geometric distribution given by Eq.~(\ref{geometric}) by the logarithmic distribution
\begin{equation}\label{logpn}
    P_K(k) = \frac{-1}{\ln(1-p)}\frac{p^k}{k}
\end{equation}
we obtain the Negative Binomial Distribution. In both case, for NBD and GPD, we have $P(0)=exp(-\lambda)$. Because the logarithmic distribution in Eq.~(\ref{logpn}) is narrower in comparison with geometric distribution Eq~(\ref{geometric}), to reproduce data on $P(n)$ the mean number of elementary emitting cells $\lambda$ should be larger. For NBD $\lambda = $ 2.55, 2.93 and 2.75 for $\sqrt{s} = $ 0.9, 2.76 and 7 TeV respectively, and are seemingly larger than $\lambda$ evaluated for GPD (see Table~\ref{tabpara}), leading for lower values of the void probability $P(0)$. For GPD we have $P(1)/P(0)=\lambda\theta < 1 $ for values from Table~\ref{tabpara}. Based on this approach, we argue that quantum-statistical interference must occur for identical photons produced in high-energy interactions. Only a small number of elementary emitting cells produced in a multiparticle production reaction, with occupancy numbers for photons in individual cells that are not small, can potentially lead to the experimentally observed distribution. This effect can be analyzed experimentally by studying Bose-Einstein correlations of the produced photons, which is the counterpart of the Hanbury-Brown-Twiss effect in optics \cite{BIYAJIMA1996297}.

It is also worth noting that the multiplicity distribution in the QCD parton cascade has a general form given by Eq.~(\ref{geometric}) where parameter $\theta=1/N$ is related to the average number of partons N \cite{PhysRevD.102.074008, PhysRevD.95.114008}.  In small $x$ and large rapidity $y$ regime, we have $\theta = exp\left[-4\ln2\bar{\alpha_S}y\right]$ where the strong coupling constant $\bar{\alpha_S} = \alpha_{S}N_c/\pi$ and $N_C$ is the numer of colors. For estimating $\theta$ parameters (given in Table~\ref{tabpara}) we have relatively large coupling constants $\bar{\alpha_S} = $ 0.12, 0.16 and 0.19 for $\sqrt{s}$= 0.9, 2.76 and 7 TeV respectively. It is possible, that we cannot consider hadrons as the dilute system of partons but rather have to consider them as the dense system of partons. For such a situation, we expect that $1/\theta = N \sim Q_S^2 $ where $Q_S$ denotes the saturation scale \cite{PhysRevD.102.074008}. Our results show power law dependence on Feynman-x variable $ Q_S^2 \sim x^{-\Lambda}$ with power index $\Lambda=0.31$. The result is quite close to the values $\Lambda =$ 0.29 - 0.30, found by an analysis of deep inelastic scattering and diffraction \cite{PhysRevD.59.014017, PhysRevD.60.114023} as well as theoretical studies of QCD dynamics \cite{TRIANTAFYLLOPOULOS2003293}. Our results can hopefully lead to wider theoretical investigations and a better understanding of photon production in ultrarelativistic hadronic collisions.
\begin{acknowledgements}
RRN acknowledges a very helpful correspondence with Professor Ding-Wei Huang (Chung Yuan Christian University, Taiwan). GW was supported in part by the Polish Ministry of Education and Science, grant Nr 2022/WK/01.
\end{acknowledgements}
%

\begin{thebibliography}{31}%

\bibitem{Abelev2015}
B. Abelev \emph{et al}. (ALICE Collaboration), \href {https://doi.org/10.1140/epjc/s10052-015-3356-2}{Eur. Phys. J. C 75, 146 (2015).}

\bibitem{Chau:1992uq}
 L. L. Chau and D.W. Huang, \href {https://doi.org/10.1140/epjc/s10052-015-3356-2}{Phys. Lett. B  283, 1 (1992).}

\bibitem{PhysRevLett.70.3380}
 L. L. Chau and D.W. Huang, \href {https://doi.org/10.1103/PhysRevLett.70.3380}{Phys. Rev. Lett. 70, 3380 (1993).}

\bibitem{Huang:1996az}
 D.W. Huang, \href {https://doi.org/10.1142/S021773239600268X}{Mod. Phys. Lett. A, Vol. 11, No. 34, pp. 2681-2692 (1996).}

\bibitem{PhysRevD.37.2446}
 D.W. Huang, \href {https://doi.org/10.1103/PhysRevD.37.2446}{Phys. Rev. D 37, 2446 (1988).}
 
 \bibitem{Huang:1997bra}
 D.W. Huang, \href {https://doi.org/10.1088/0954-3899/23/8/004}{J. Phys. G: Nucl. Part. Phys. 23 895 (1997)}

\bibitem{PhysRevD.10.65}
 W. J. Knox, \href {https://doi.org/10.1103/PhysRevD.10.65}{Phys. Rev. D 10, 65 (1974).}

\bibitem{Kauffmann_1978}
S. K. Kauffmann and M. Gyulassy, \href {https://doi.org/10.1088/0305-4470/11/9/006}{J. Phys. A: Math. Gen. 11 1715 (1978)}

\bibitem{Vasudevan_1984}
R. Vasudevan, P. R. Vittal, and K. V. Parthasarathy, \href {https://doi.org/10.1088/0305-4470/17/5/022}{J. Phys. A: Math. Gen. 17 989 (1984)}

\bibitem{BALANTEKIN1991231}
A.B. Balantekin, J.E. Seger \href {https://doi.org/https://doi.org/10.1016/0370-2693(91)91031-P}{Phys. Lett. B  266, 3-4 (1991).}

\bibitem{HEGYI1999126}
S. Hegyi, \href {https://doi.org/https://doi.org/10.1016/S0370-2693(99)00957-0}{Phys. Lett. B 463, 126 (1999).}

\bibitem{Wilk_2017}
G. Wilk and Z. W\l odarczyk \href {https://doi.org/https://doi.org/10.1088/0954-3899/44/1/015002}{J. Phys. G: Nucl. Part. Phys. 44 015002 (2017).}
 
\bibitem{Ang2020}
H. W. Ang, A. H. Chan, M. Ghaffar, M. Rybczy\'{n}ski, G. Wilk and Z. W\l odarczyk \href {https://doi.org/10.1140/epja/s10050-020-00140-w}{Eur. Phys. J. A 56, 117 (2020)}

\bibitem{PhysRevD.99.094045}
M. Rybczy\'{n}ski, G. Wilk and Z. W\l odarczyk \href {https://doi.org/https://doi.org/10.1103/PhysRevD.99.094045}{Phys.
Rev. D 99, 094045 (2019).}

\bibitem{EPJC(2018)78:816} 
I. Zborovsk\'{y} \href{https://doi.org/10.1140/epjc/s10052-018-6287-x}{Eur. Phys. J. C (2018) 78:816}


 
\bibitem{Collaboration_2008}
K. Aamodt \emph{et al}. (ALICE Collaboration), \href {https://doi.org/10.1088/1748-0221/3/08/S08002}{JINST 3 S08002 (2008).}

\bibitem{Aamodt_2010}
K. Aamodt \emph{et al}. (ALICE Collaboration), \href {https://doi.org/10.1140/epjc/s10052-010-1339-x}{Eur. Phys. J. C 68, 89-108 (2010)}

\bibitem{ALICE:PMD1}
G. Dellacasa \emph{et al}. (ALICE Collaboration), \href{http://cdsweb.cern.ch/search?ln=en&amp;sysno=002399082CER}{CERN-LHCC-99-32, CERN-OPEN-2000-184, ALICE-TDR-6 (1999)}
 
\bibitem{ALICE:PMD2}
ALICE Collaboration, \href{http://cdsweb.cern.ch/search?ln=en&amp;sysno=002399082CER}{CERN-LHCC-2003-038; ALICE-TDR-6-add-1 (2003)}

\bibitem{AIHP_1930__1_2_117_0}
G. P\'olya , \href{http://archive.numdam.org/item/AIHP_1930__1_2_117_0/}{Annales de l'institut Henri Poincar\'e, Volume 1 (1930) no. 2, pp. 117-161}
 
\bibitem{Aeppli}
G. P\'olya , \href{https://www.research-collection.ethz.ch/bitstream/handle/20.500.11850/132141/eth-20128-01.pdf}{Leemann \& Co. A.-G. (1924)}, On the theory of chained
  probabilities: Higher-order Markov chains (German).

\bibitem{https://doi.org/10.1002/nav.3800150206}
C. C. Sherbrooke, \href{https://doi.org/https://doi.org/10.1002/nav.3800150206}{Naval Research Logistics, 15: 189-203 (1968).} 

\bibitem{doi:10.1080/00949650802711925}
G. \"Ozel and C. \.Inal \href{https://doi.org/10.1080/00949650802711925}{ Journal of Statistical Computation and Simulation, 80:5, 479-487 (2010)} 

\bibitem{EVANS1953}
D. A. Evans,  \href{https://www.tandfonline.com/doi/abs/10.1080/03461238.1960.10410586}{ Biometrika 40, 186 (1953).} 
  
\bibitem{doi:10.1080/03461238.1960.10410586}
C. Philipson, \href{doi:10.1080/03461238.1960.10410586}{Scandinavian Actuarial Journal, 1960:3-4, 136-162.}

\bibitem{doi:10.1287/opre.7.3.362}
H. P. Galliher, P. M. Morse and M. Simond, \href{https://doi.org/10.1287/opre.7.3.362}{Operations Research 1959 7:3, 362-384}
   
\bibitem{BIYAJIMA1996297}
M. Biyajima, N. Suzuki, , G. Wilk and Z. W\l odarczyk \href{https://doi.org/https://doi.org/10.1016/0370-2693(96)00894-5}{Phys. Lett. B  386, 297 (1996)}

\bibitem{PhysRevD.102.074008}
E. Gotsman and E. Levin, \href{https://doi.org/10.1103/PhysRevD.102.074008}{Phys. Rev. D 102,
074008 (2020)}

\bibitem{PhysRevD.95.114008}
D. E. Kharzeev and E. M. Levin,  \href{https://doi.org/10.1103/PhysRevD.95.114008}{Phys. Rev. D 95,
114008 (2017).}

\bibitem{PhysRevD.59.014017}
K. Golec-Biernat and M. W\"usthoff,  \href{https://doi.org/10.1103/PhysRevD.59.014017}{Phys. Rev. D
59, 014017 (1998).}

\bibitem{PhysRevD.60.114023}
K. Golec-Biernat and M. W\"usthoff,  \href{https://doi.org/10.1103/PhysRevD.60.114023}{Phys. Rev. D
60, 114023 (1999).}

\bibitem{TRIANTAFYLLOPOULOS2003293}
D.N. Triantafyllopoulos \href{https://doi.org/https://doi.org/10.1016/S0550-3213(02)01000-3}{Nucl. Phys. B. 648, 293 (2003)}
 
\end{thebibliography}
\end{document}